\documentclass[10pt, conference, letterpaper]{IEEEtran}
\IEEEoverridecommandlockouts
\usepackage{cite}
\usepackage{amsmath,amssymb,amsfonts}
\usepackage{algorithmic}
\usepackage{graphicx}
\usepackage{textcomp}
\usepackage{makecell}
\usepackage{array}
\usepackage{xcolor}
\usepackage{subcaption}
\usepackage{todonotes}
\usepackage{multirow}
\usepackage{placeins}
\usepackage[
  colorlinks=true,
  citecolor=blue,
  urlcolor=black,
  linkcolor=black
]{hyperref}\usepackage{url}
\usepackage{dblfloatfix} 
\usepackage{tikz}

\def\checkmark{\tikz\fill[scale=0.4](0,.35) -- (.25,0) -- (1,.7) -- (.25,.15) -- cycle;} 

\usepackage[
  letterpaper, 
  top=1.0in, 
  bottom=1.0in, 
  left=0.75in, 
  right=0.75in
]{geometry}

\begin{document}

\title{Hardware Generation and Exploration  of Lookup Table-Based Accelerators for 1.58-bit LLM Inference}

\author{
    \IEEEauthorblockN{Robin Geens$^*$, Joran Heldens$^*$, Joren Dumoulin, and Marian Verhelst}
    \IEEEauthorblockA{MICAS, KU Leuven, Leuven, Belgium \\
    robin.geens@kuleuven.be}
    \thanks{$^*$Equal contribution.}
}

\maketitle

\begin{abstract}

Ternary weight quantization (e.g., BitNet b1.58) offers a promising path to mitigate the memory bandwidth bottleneck in Large Language Model (LLM) inference. However, conventional compute platforms lack native support for ternary-weight arithmetic, often relying on inefficient dequantization. Lookup table (LUT)-based hardware architectures provide an effective alternative by replacing multiplications with conditional additions, but their design space remains largely unexplored. Existing designs rely on heuristic parameter selection, lacking a systematic understanding of the architectural trade-offs.

This work addresses this gap by formalizing the design space of ternary LUT-based accelerators and presenting an open-source hardware generator coupled with an analytical cost model, validated against synthesis in TSMC 16nm technology.
By spanning the full architectural space, this framework not only enables rapid design space exploration but also establishes a common footing for fair cross-design evaluation, which was previously hindered by inconsistent instantiations across published accelerators.

Using this framework, we challenge several assumptions and design choices in recent literature. We demonstrate that the optimal architecture is fundamentally governed by the activation data type: while LUT-based reuse offers significant gains for high-cost arithmetic (e.g., FP16), it yields diminishing returns for small integer types. Furthermore, we show that maximizing core size consistently improves area density compared to highly tiled approaches. Our optimized designs achieve a $2.2\times$ area reduction compared to multiplier-based baselines. Moreover, by benchmarking state-of-the-art implementations against our model, we reveal that correcting suboptimal parameters yields up to a $1.2\times$ area improvement. These findings underscore that achieving state-of-the-art efficiency requires moving beyond heuristics to a model-driven design methodology. This transition is enabled by our open-source hardware generator available at \textit{\url{https://github.com/KULeuven-MICAS/ternary-lut-dse}}.

\end{abstract}
\section{Introduction}

Transformer-based Large Language Models (LLMs) have achieved breakthrough performance across a wide range of natural language processing tasks. This success has been enabled by rapid scaling in model size, but scaling also increases the computational and memory demands of inference. 

In particular, the auto-regressive decode stage becomes increasingly memory-bound: unlike the prefill stage, where weights are reused across many tokens, decode processes only one token at a time. Consequently, the cost of fetching weights dominates, leading to underutilized compute hardware and limiting throughput by off-chip bandwidth.

Extreme weight-only quantization formats, such as BitNet~\cite{wang_bitnet_2023} and BitNet b1.58~\cite{ma2024era1bitllmslarge}, address this bottleneck by aggressively reducing weight precision to ternary values $\{-1, 0, +1\}$. This significantly reduces bandwidth requirements and, in theory, lowers arithmetic complexity. However, mainstream compute platforms (CPUs, GPUs) lack native support for mixed-precision ternary arithmetic. Instead, they rely on dequantization and convert ternary weights to higher precision (e.g., FP16) before multiplication, eliminating the potential compute density and energy gains of the format~\cite{park_lut-gemm_2024}. To fully exploit these efficient quantization schemes, specialized hardware architectures are required.

\begin{figure}[t]
    \centerline{\includegraphics[width=0.5\textwidth]{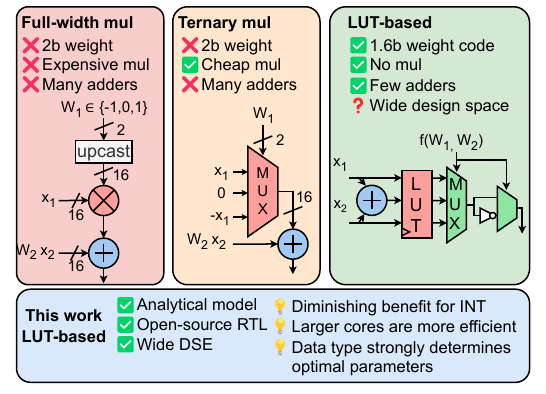}}
    \caption{Processing elements for ternary weight multiplication.    }
    \label{fig:paper-overview}
\end{figure}

Lookup Table (LUT)-based architectures offer a viable solution for such low-cardinality weight computations. As illustrated in Figure~\ref{fig:paper-overview}, conventional architectures waste resources on complex high-precision Multiply-and-Accumulate (MAC) units. A naive adaptation (center) might use mixed-precision multipliers, but this still incurs overhead. LUT-based approaches (right) replace multiplications entirely with conditional additions, enabling both spatial and temporal reuse of intermediate partial sums. While prior works have explored LUTs in both hardware and software~\cite{jeon_biqgemm_2020, park_lut-gemm_2024, park_figlut_2025, qiao_tellme_2025, kim_slim_lama}, they typically present isolated design points optimized for specific throughput targets. It remains unclear whether these implementations are truly optimal, coincidentally effective, or sub-optimal compared to naive baselines. Furthermore, systematic exploration of the trade-offs between LUT group size, tile geometry, and activation data type is notably absent in the literature.

To address this gap, we formalize the design space of LUT-based architectures and present an open-source hardware generator capable of producing RTL for any parameter instantiation within this space. This enables the first systematic design-space exploration (DSE) of ternary accelerators, providing analytical and empirical insights into area and scalability. The generator also offers a common footing for cross-instantiation comparison, addressing the difficulty of evaluating prior designs implemented at incompatible operating points. Using this unified framework, we contextualize existing architectures and extract practical guidelines to inform future ternary-weight inference accelerator design.

The key contributions of this work are:
\begin{itemize}
    \item \textbf{Hardware generator} (Section~\ref{sec:hw_arch}): We formalize the design space of ternary LUT-based matrix multiplication accelerators and propose an open-source hardware generator in Chisel to instantiate any point within this design space.
    \item \textbf{Companion analytical model} (Section~\ref{sec:cost_model}): We develop an analytical area and throughput model that captures the behavior of our hardware generator. Validated against TSMC 16nm synthesis results for both INT and FP activation types (Section~\ref{sec:validate}), this model enables rapid exploration without requiring exhaustive synthesis.
    \item \textbf{Exploration and insights} (Section~\ref{sec:dse}-\ref{sec:results}): We conduct a comprehensive DSE, compare our LUT-based designs to multiplier-based baselines and state-of-the-art works, and provide concrete guidelines for efficient accelerator design.
\end{itemize}

\section{Background} \label{sec:background}
\subsection{Weight-only Ternary Quantization}
While early quantization schemes focused on INT8 or INT4 precision, recent extreme weight-only quantization methods have pushed precision down to ternary values. As summarized in Table~\ref{tab:quant_schemes}, schemes such as BitNet~\cite{wang_bitnet_2023}, BitNet b1.58~\cite{ma2024era1bitllmslarge}, and follow-up ternary formulations ~\cite{JMLR:v26:24-2050,chen2024,kaushal2024spectrasurprisingeffectivenesspretraining,sundaram2024llavaolmobitnet1bternaryllmgoes} demonstrate that LLMs can maintain competitive accuracy with only $\log_2(3) \approx 1.58$ bits per weight thanks to quantization-aware training. Notably, their accuracy gap relative to full precision decreases for larger models~\cite{badshah2024quantifyingcapabilitiesllmsscale}, making ternary formats particularly attractive for modern LLM scales.

This reduction offers two distinct advantages for hardware. First, it minimizes the memory bandwidth required during the memory-bound decode stage. Second, it simplifies arithmetic: ternary multiplication reduces to conditional addition (accumulation of $+x$, $-x$, or $0$). However, standard CPUs, GPUs, and TPUs must dequantize these weights to higher precision to use their existing MAC units, negating the computational/area benefits~\cite{park_lut-gemm_2024}. Native hardware support is therefore required to translate the theoretical sparsity and simplicity of ternary weights into actual energy and area savings.

\begin{table}[h]
\centering
\caption{Current Binary and Ternary LLMs}
\label{tab:quant_schemes}
\begin{tabular}{lllll}
\hline
Model        & \begin{tabular}[c]{@{}c@{}}Weights\\ Precision\end{tabular} & \begin{tabular}[c]{@{}c@{}}Activation\\ Precision\end{tabular} & \begin{tabular}[c]{@{}c@{}}Accuracy \\ Loss\end{tabular} &  \\ \hline
BitNet~\cite{wang_bitnet_2023}      & 1 bit & INT8 & -3.3\% &  \\
BitNet b1.58~\cite{ma2024era1bitllmslarge} & 1.58 bit & INT8 & +1\% &  \\
BitNet V2~\cite{wang2025bitnetv2native4bit}    & 1 bit & INT4 & &  \\
TernaryLLM~\cite{chen2024}   & 1.58 bit & FP16 & -6.2\% &  \\ \hline
\end{tabular}
\end{table}

\subsection{Lookup Table-Based Multiplication}

A practical method to directly support low-bit computations is lookup table (LUT)-based multiplication. The fundamental working principle is that, because a low-precision weight matrix contains only a small number of distinct values, many of the required multiplications in a matrix–vector product are repeated. These repeated partial results can be precomputed once, stored in LUTs, and then reused through indexed lookups.

\begin{figure}[h!]
    \centerline{\includegraphics[width=0.45\textwidth]{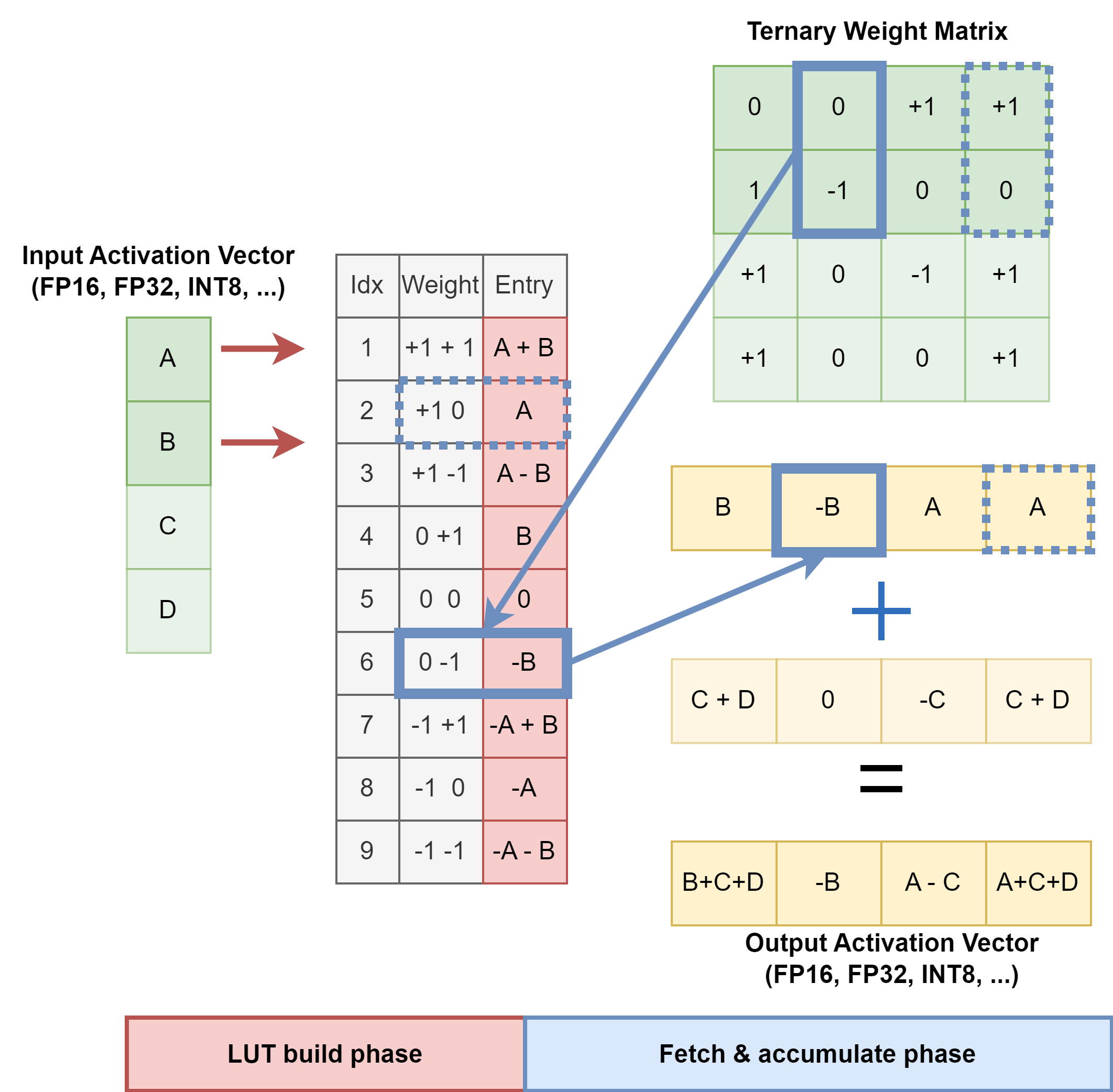}}
    \caption{Principle of LUT-based GEMV multiplication,
    illustrating the two computation phases ($\mu=2$).}
    \vspace{-3mm}
    \label{fig:lut_gemv}
\end{figure}

As shown in Figure~\ref{fig:lut_gemv}, the operation is split into two phases. In the \textbf{LUT Build Phase}, all $3^\mu$ ternary combinations of $\mu$ input activations are computed and stored in LUTs, where $\mu$ is a design parameter. In the \textbf{Fetch \& Accumulation Phase}, groups of $\mu$ ternary weights act as indices to retrieve the corresponding precomputed partial sums from the LUTs. The fetched values are accumulated to form the final output. The LUT entries can be reused multiple times before they have to be replaced in a new LUT Build Phase.

\subsection{Lookup-Table Based Architecture}

\begin{figure}[tb]
    \centerline{\includegraphics[width=\linewidth]{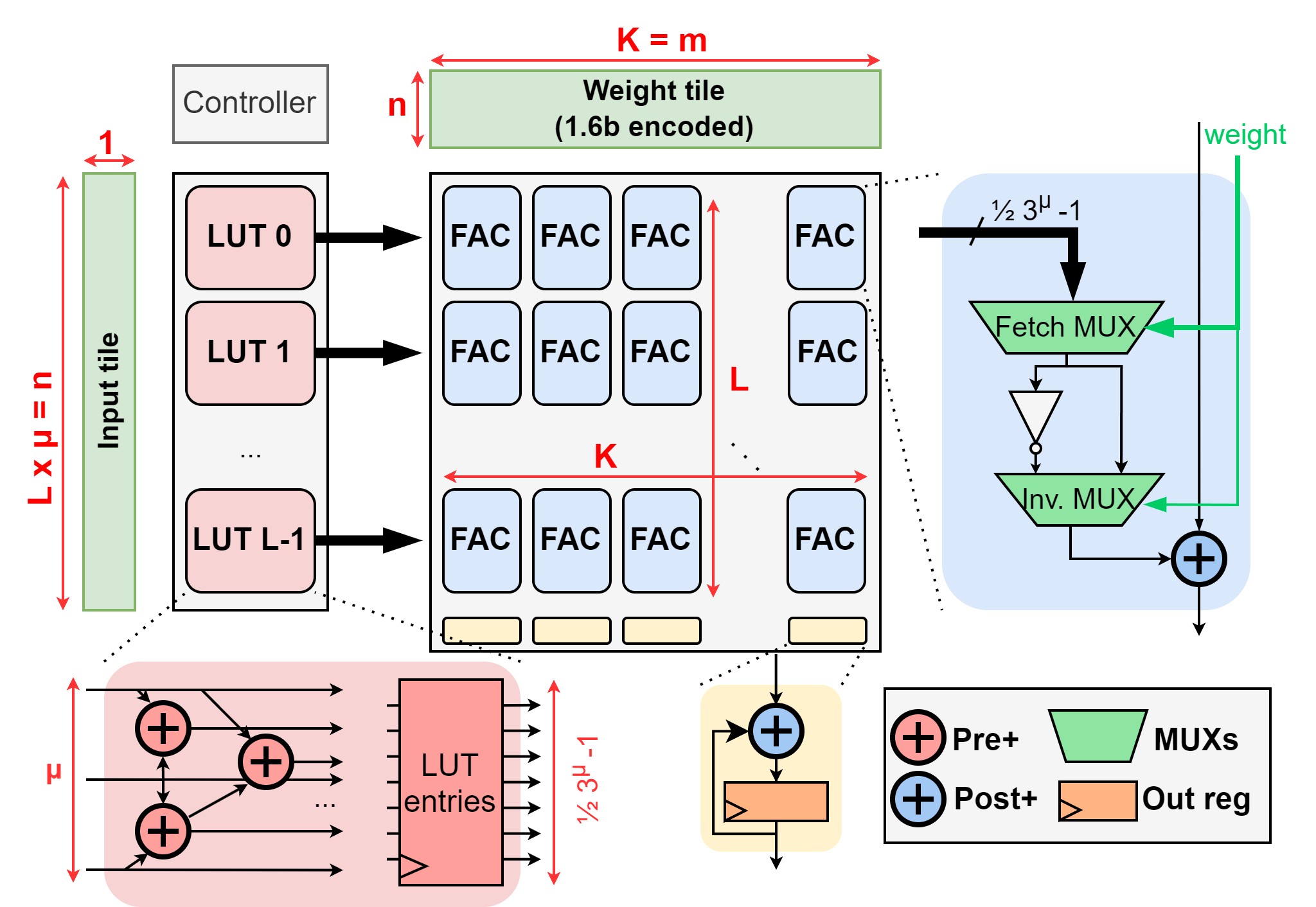}}
    \caption{Block diagram of the LUT-based ternary matrix multiplication architecture. Key submodules are categorized as Pre+, Post+, MUXs or Output Registers.
    }
    \label{fig:arch_overview}
    \vspace{-3mm}
\end{figure}

The conceptual phases of Figure~\ref{fig:lut_gemv} can be computed efficiently on a custom hardware architecture, as shown in Figure~\ref{fig:arch_overview}. 
The \textbf{LUT Build Phase} is implemented using an adder-tree structure that generates all ternary combinations of $\mu$ activations, followed by a register file that stores the resulting LUT entries.
The \textbf{Fetch \& Accumulation Phase} is realized by read-out multiplexers that select LUT entries based on $\mu$-weight groups. The selected values are accumulated over time in an output accumulator.

To increase throughput and efficiency, spatial parallelism can be exposed along two orthogonal dimensions. First, $L$ LUTs can be instantiated in parallel, each operating on a distinct group of $\mu$ activations, enabling spatial weight reuse. Their outputs are accumulated spatially before entering the output accumulator. Second, $K$ parallel read-out MUXes per LUT enable multiple simultaneous index operations on the same LUT entries.

The architecture is defined by four design parameters:
{\boldmath
\begin{enumerate}
  \item $\mu$: LUT group size, determining LUT depth
  \item $L$: number of parallel LUTs
  \item $K$: number of parallel fetchers per LUT
  \item The activation data type (e.g., INT8, FP16) 
\end{enumerate}
}

While this work focuses on GEMV operations, which dominate the LLM decode stage, the design naturally extends to GEMM. Because the core architecture is optimized for maximum compute-density per unit area, the fundamental throughput engine remains invariant even when weights are reused across a batch. The GEMV operation is executed over multiple cycles, each processing a tile of the ternary matrix with size

\begin{equation}
    \text{tile size} = n \times m = (L \cdot \mu) \times K
\end{equation}

\noindent The tile size directly determines the accelerator's peak throughput as $f_{clk} \cdot n \cdot m$.

\subsection{Prior Works}
\noindent \textbf{BiQGEMM (2020)~\cite{jeon_biqgemm_2020}} was one of the earliest works to introduce LUT-based GEMM for binary and few-bit quantization levels. It formalized analytical expressions to guide the design, including strategies like dynamic programming to minimize redundant computation in the precompute phase. However, its application was restricted to CPUs and GPUs.

\noindent \textbf{LUT-GEMM (2024)~\cite{park_lut-gemm_2024}} extended this concept to support multiple quantization schemes, including both uniform and non-uniform formats. Its focus was on mitigating the communication overhead between GPUs when deploying large LLMs. Like BiQGEMM, it remained primarily software-based.

\noindent \textbf{LUT Tensor Core (2025)~\cite{mo_lut_2025}} introduced a co-design approach, combining GPU-based precomputation with hardware-supported fetching and accumulation from LUTs. 

\noindent \textbf{FIGLUT (2025)~\cite{park_figlut_2025}} introduced a fully specialized binary LUT hardware architecture that supports multiple quantization levels using bit-serial techniques.

\noindent \textbf{TeLLMe v2 (2025)~\cite{tellmev2}} proposed the first end-to-end FPGA accelerator for ternary LLMs at the edge.

\noindent \textbf{TENET (2025)~\cite{tenet}} designed a LUT-centric architecture for LLM inference at the edge, complementing the LUT-based cores with higher-precision attention cores.

\noindent \textbf{Slim-LLaMa (2025)~\cite{kim_slim_lama}} designed and implemented a fully binary and ternary LUT-based inference accelerator, fabricated in a 28nm process node. By natively supporting ternary quantization, this silicon-proven implementation can run LLM tasks at 189.8 TOPS/W.

\begin{table}[h]
\centering
\renewcommand{\arraystretch}{1.5}
\caption{Summary of related works.}
\begin{tabular}{l c c c c c c }
\hline
Work & Impl. & Ternary & \makecell{Act.\\type} & $L$ & $\mu$ & $K$ \\ 
\hline
BiQGEMM~\cite{jeon_biqgemm_2020}   & SW  & $\times$  & FP16  & N/A & N/A & N/A \\
LUT-GEMM~\cite{park_lut-gemm_2024} & SW  & $\times$  & FP16  & N/A & N/A & N/A \\
\makecell[l]{LUT Tensor\\Core~\cite{mo_lut_2025}} & \makecell{co-\\design} & $\times$  & \makecell{INT8/16\\FP8/16}  & N/A & N/A & N/A \\
FIGLUT~\cite{park_figlut_2025}  & HW & $\times$  & \makecell{FP16/32\\BF16}  & 32 & 4 & 32 \\ 
Slim-Llama~\cite{kim_slim_lama} & HW & \checkmark  & INT8 &  8 & 3 &  2 \\ 
TENET~\cite{tenet}             & HW & \checkmark  & INT8 & 16 & 2 & 64 \\ 
TellMe-v2~\cite{tellmev2}      & HW & \checkmark  & INT8 & 28 & 3 & 23 \\ 
\hline
\textbf{This}  & \textbf{HW}  & \textbf{\checkmark }  & \makecell{Any\\INT/FP} 
 & Any & Any & Any \\
\hline
\end{tabular}
\label{tab:comparison_work}
\end{table}

Table \ref{tab:comparison_work} summarizes prior works. While these efforts demonstrate the viability of LUT-based acceleration, they remain narrow in scope. 

Most focus on binary or power-of-two quantization, evaluate only a single operating point, or lack a systematic analysis of architectural trade-offs. Across the literature, hardware parameters vary widely and are often selected heuristically, leaving it unclear whether these choices are optimal. To date, no work has systematically mapped the design space
to determine which parameter sets are optimal for a given constraint, leaving a gap that this work aims to fill.

\section{Hardware Generator Design} \label{sec:hw_arch}
To assess the impact of parameter choices, we developed a fully parametrizable RTL generator for LUT-based arithmetic. This generator is written in Chisel, an object-oriented hardware construction language, and utilizes open-source floating-point units~\cite{chisel-float} to support diverse data types. The generator is parameterized to cover the full design space ($L$, $\mu$, $K$, data type), providing a consistent baseline for comparing distinct architectural configurations.

\subsection{Architectural Overview}
Figure~\ref{fig:arch_overview} presents the representative hardware architecture, with the key parameters $L$, $\mu$, $K$ highlighted in red. It consists of three main components: 

\begin{itemize}
    \item \textbf{LUTs:} Responsible for the \textit{Build Phase}. They contain an adder network to generate LUT entries from input activations and registers to store them. This network is fully pipelined to support high-frequency operation.
    \item \textbf{Fetch-And-Accumulate (FAC) Units:} Responsible for the \textit{Fetch and Accumulate Phase}. They select LUT entries based on indices derived from the weights and accumulate the results from $L$ parallel LUTs.
    \item \textbf{Output Buffers:} Store and combine partial sums across cycles to maintain output stationarity, allowing the accumulation of results over the full matrix dimension.
\end{itemize}

\subsection{LUT Module Optimization}
The accelerator comprises $L$ parallel lookup tables, each storing the partial sums of $\mu$ activations.
A naive implementation for group size $\mu$ would store $3^\mu$ entries, computed with $(\mu - 1) \cdot (3^{\mu} - 1)$ additions. To minimize the area overhead of the adder tree and storage registers, we employ three optimization techniques that reduce both the number of LUT entries and the required number of additions:

\begin{enumerate}
    \item \textbf{Symmetry Reduction:} Since the ternary set is symmetric around zero, only the "positive" half of the combinations needs to be stored~\cite{mo_lut_2025}. The negative counterparts are derived dynamically by inverting the stored entry. For instance, the result for weights $\{-1, +1\}$ is obtained by negating the stored entry for $\{+1, -1\}$. This halves the required storage and pre-computation adders.
    \item \textbf{Redundancy Elimination:} The adder tree is structured to maximize the reuse of intermediate sums~\cite{jeon_biqgemm_2020}. For example, the entries for $\{+1, +1, 0\}$ and $\{+1, +1, +1\}$ share the sub-expression $\{+1, +1\}$, allowing a single adder to serve multiple outputs.
    \item \textbf{Sparsity:} Ternary weights include zero. Additions involving zero-weighted inputs are identity operations and are pruned from the adder tree entirely.
\end{enumerate}

\begin{figure}[tb]
    \centerline{\includegraphics[width=0.8\linewidth]{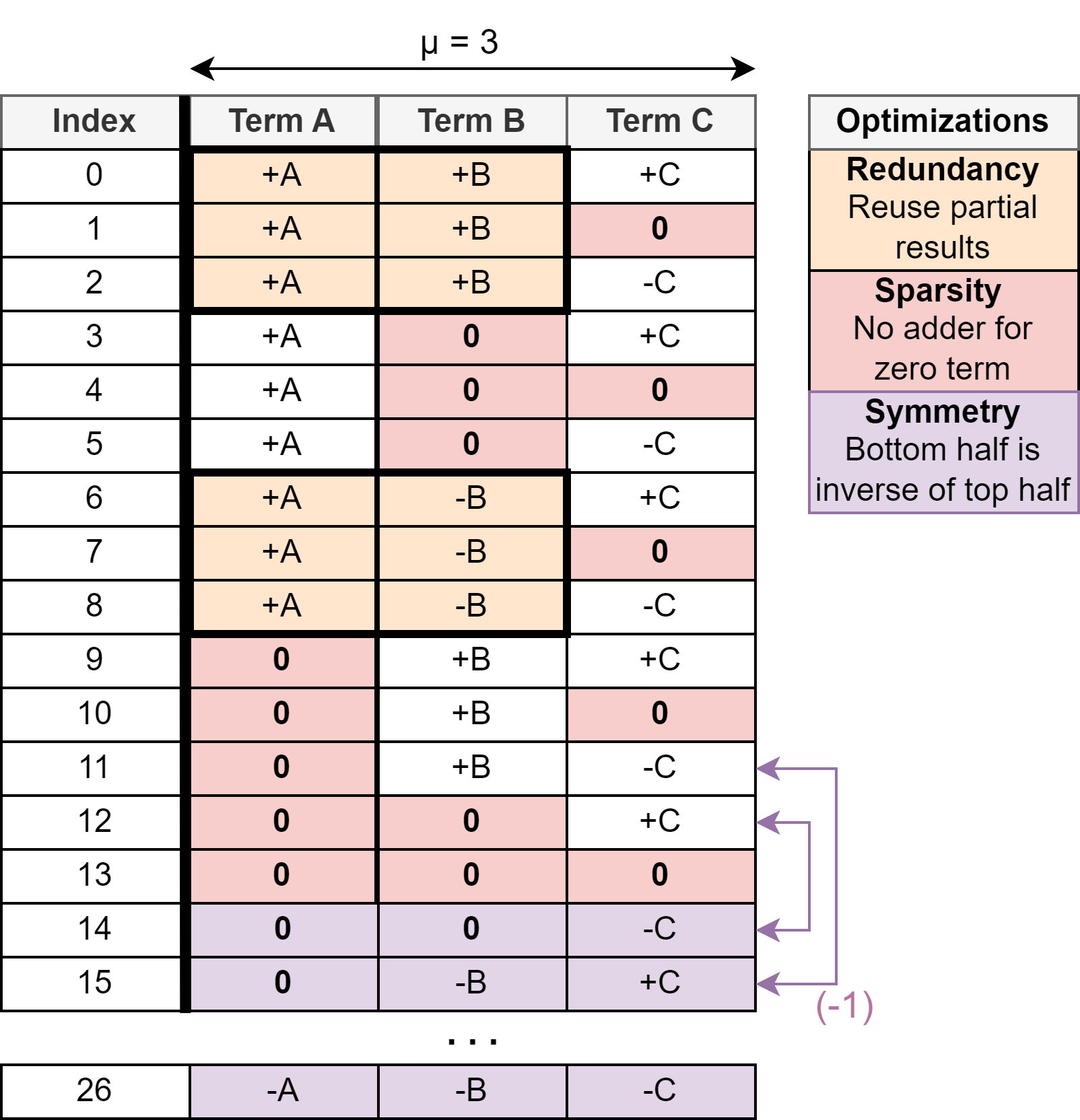}}
    \caption{Optimizations to reduce the number of LUT entries and required adders to compute the entries. Symmetry, Redundancy, and Sparsity reductions are applied explicitly in the generator.
    }
    \label{fig:precompute_table_opts}
\end{figure}

\noindent Figure~\ref{fig:precompute_table_opts} illustrates these techniques. The first optimization halves both the number of entries and required additions, while the second and third only reduce the number of required adders in the LUT Build Phase adder tree. Although a subset of these optimizations can also be found by combinatorial optimization tools, we have found through experimentation that this is not the case when the adder tree structure is heavily pipelined. In this case, explicitly programming the optimization in RTL is required.

The number of adders required per LUT after optimization is bounded by:

\begin{equation}
\frac{ \text{\#adders}}{\text{LUT}} \leq (\mu - 1) \cdot \frac{3^{\mu} - 1}{2} - R(\mu) - \mu \cdot S(\mu)
\label{eq:additions_per_lut}
\end{equation}

\begin{equation}
\begin{cases}
S(\mu) = S(\mu - 1) + 3^{\mu - 2}, & \mu > 2 \\
S(2) = 1
\end{cases}
\label{eq:adders_sparsity}
\end{equation}

\begin{equation}
R(\mu) =2 \sum_{k=0}^{\mu - 3} 2^k \cdot (3^{\mu-2-k} - 1)
\label{eq:adders_redundancy}
\end{equation}

Here, $R(\mu)$ and $S(\mu)$ capture the savings from sparsity and redundancy, respectively. For $\mu = 4$, these optimizations collectively reduce the number of required adders by as much as 81.89\% compared to a naive implementation.

\subsection{Fetch-and-Accumulate Module}

The FAC module executes the retrieval and summation of LUT entries. Each FAC unit contains a multiplexer (MUX) that selects one entry from the LUT's storage. The number of inputs to this MUX is equal to the size of the LUT table $\frac{3^\mu-1}{2}$. The selection is driven by the encoded weight key. Following the symmetry reduction technique, if the symmetry bit of the key indicates a negative combination, the MUX output is inverted (sign-flipped) before accumulation. The outputs from $L$ such FAC units are then summed via a reduction adder tree to produce a single partial sum for the output channel.

\subsection{Offline Weight Encoding}

To avoid the inefficiency of standard binary representations, we employ a dense offline encoding scheme. A group of $\mu$ ternary weights is mapped to a single index of width $\lceil \log_{2}(\frac{3^\mu - 1}{2}) \rceil + 1$. The most significant bit serves as the symmetry flag (indicating inversion), while the remaining bits drive the MUX selection. This encoding achieves an average density of $\approx 1.6$ bits per weight, within 1\% of the theoretical information-theoretic limit ($\log_2 (3) \approx 1.58$ bits). Compared to a naive 2-bit representation, this reduces memory bandwidth and on-chip storage requirements by 20\%, a critical saving for bandwidth-bound inference. As this encoding is performed offline, it incurs no runtime latency penalty.

\section{Companion cost model}\label{sec:cost_model}
In this section, we develop an analytical model to estimate the area cost of the LUT-based architecture. We subdivide the hardware design into submodules and derive scaling equations for each module as a function of key design parameters. These closed-form estimates guide design space exploration, avoiding costly synthesis for each design point. 

\subsection{Scaling Formulas of Hardware Submodules}
The following formulas describe the scaling behavior of the four major submodules defined in Figure~\ref{fig:arch_overview}:

\begin{equation}
    \text{Build+}_\text{cost} \sim \frac{\text{\#adders}}{\text{LUT}} \cdot (\# \text{LUTs}) \approx \frac{3.069^\mu}{1.938} \cdot \frac{n}{\mu}
    \label{eq:scale_pre}
\end{equation}
\begin{equation}
    \text{Accumulate+}_\text{cost} \sim L \cdot K = \frac{n \cdot m}{\mu}
     \label{eq:scale_post}
\end{equation}
\begin{equation}
    \text{MUX}_\text{cost} \sim \frac{n\cdot m}{\mu} \cdot  \frac{3^\mu-1}{2}
    \label{eq:scale_mux}
\end{equation}
\begin{equation}
    \text{OutReg}_\text{cost} \sim K =m
    \label{eq:scale_reg}
\end{equation}

$\text{Build+}_{cost}$ and $\text{Accumulate+}_{cost}$ denote the number of adders required in each respective phase. The term $\frac{\#\text{adders}}{\text{LUT}}$ is a simplified expression obtained by curve-fitting Equation~\ref{eq:additions_per_lut}. $\text{MUX}_{cost}$ represents the equivalent number of 2-to-1 multiplexers required, while $\text{OutReg}_{cost}$ corresponds to the number of word-sized accumulator registers.

For a fixed tile size ($n \cdot m$), and therefore fixed throughput, these equations reveal a fundamental trade-off between LUT Build and Accumulate hardware. Small group sizes ($\mu$) result in longer accumulation adder chains, while larger group sizes shift the cost toward building (computing the LUT's entries) due to the exponential LUT growth. An optimal $\mu$ must therefore balance these opposing contributions.

\subsection{Analytical Hardware Cost Model}
The scaling equations provide only qualitative insight into how the architecture grows with group size and tile dimensions. To quantitatively estimate the required area for a given set of parameters and to identify efficient design points more precisely, the total area must be expressed as a weighted sum of the main contributors:

\vspace{-2mm}
\begin{align}
\text{Area} 
   &= a_{add} \cdot \left( \text{Build+}_\text{cost} + \text{Accumulate+}_\text{cost} \right) \notag \\
   &\quad + (a_{mux} + a_{inv}) \cdot \text{MUX}_\text{cost} 
          + a_{reg} \cdot \text{OutReg}_\text{cost}
    \label{eq:total_hw_cost}
\end{align}

\noindent Here, the coefficients $a_{x}$ represent the technology-dependent area of the fundamental unit cells:
    \begin{itemize}
    \item $a_{add}$: Area of a single scalar adder of chosen activation type.
    \item $a_{mux}$: Area of a word-sized 2-to-1 multiplexer.
    \item $a_{inv}$: Overhead of the sign-inversion logic for one activation value.
    \item $a_{reg}$: Area of a word-sized register.
\end{itemize}

\noindent These coefficients allow the model to adapt to different activation data types. By keeping these coefficients as variables, the model supports rapid sensitivity analysis across technologies and precisions.

\section{Model Validation And Analysis}\label{sec:validate}
In this section, we evaluate the performance of our hardware generator and validate the analytical model by synthesizing a multitude of designs in TSMC 16nm technology using Cadence Genus. The implementation is fully pipelined to reliably meet 500~MHz timing constraints at every design point.

\begin{table}[tbh]
    \caption{Explored design space for model evaluation.}
    \centering
    \renewcommand{\arraystretch}{1.2} 
    \begin{tabular}{l c}
    \hline
    \textbf{Parameter} & \textbf{Values} \\
    \hline
    Tile size ($n = m$) & 8, 32, 64, 96 \\
    Group size ($\mu$) & 1, 2, 3, 4, 5 \\
    Activation (adder) data type & FP16, INT8 \\
    \hline
    \end{tabular}
    \label{tab:design_space}
\end{table}

\begin{figure}[tb]
    \centering
    \begin{subfigure}[b]{0.45\textwidth}
        \includegraphics[width=\textwidth]{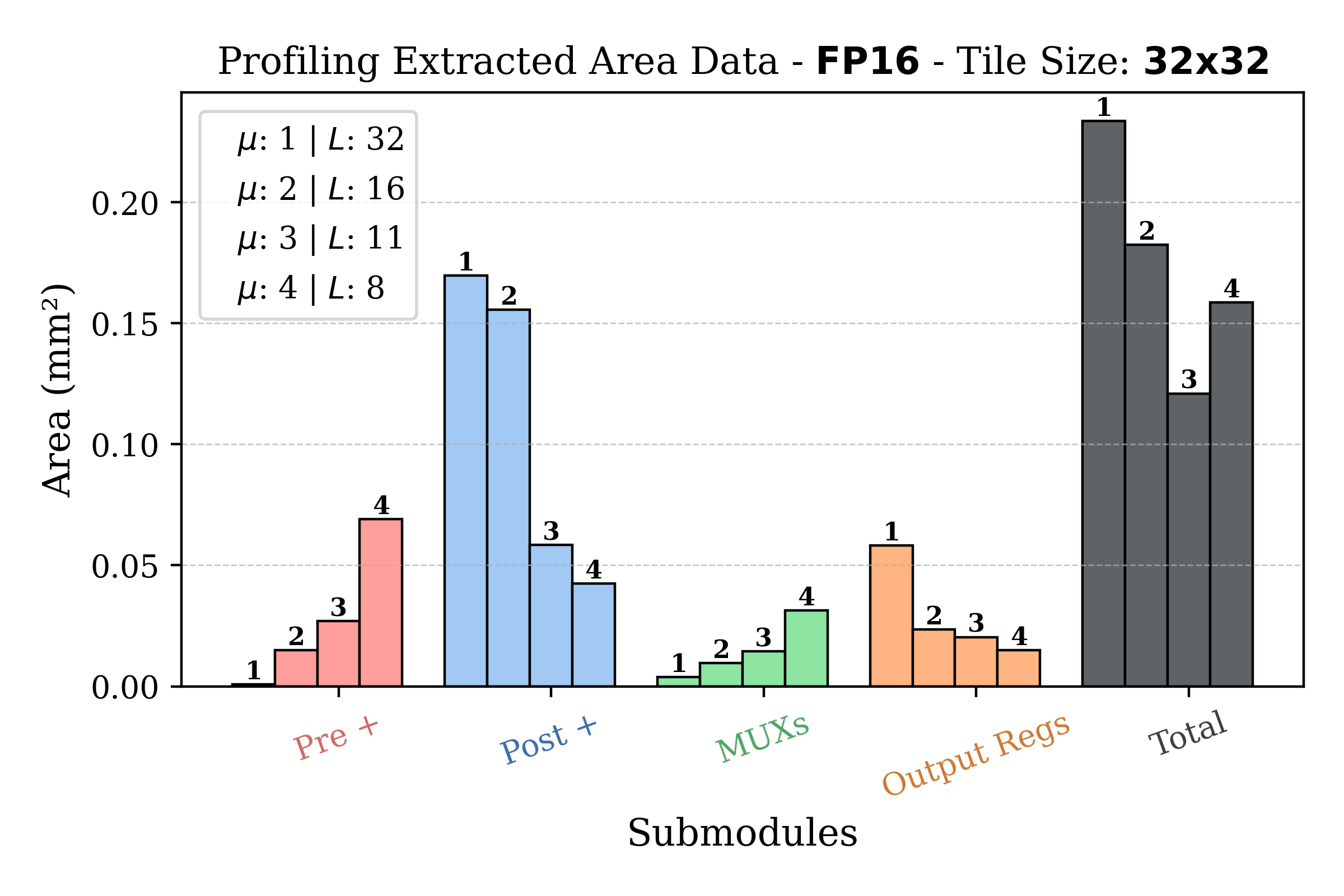}
        \vspace{-4mm}
        \caption{Synthesis area results.}
        \label{fig:surface_plot_int}
    \end{subfigure}
    \hfill
    \begin{subfigure}[b]{0.45\textwidth}
        \includegraphics[width=\textwidth]{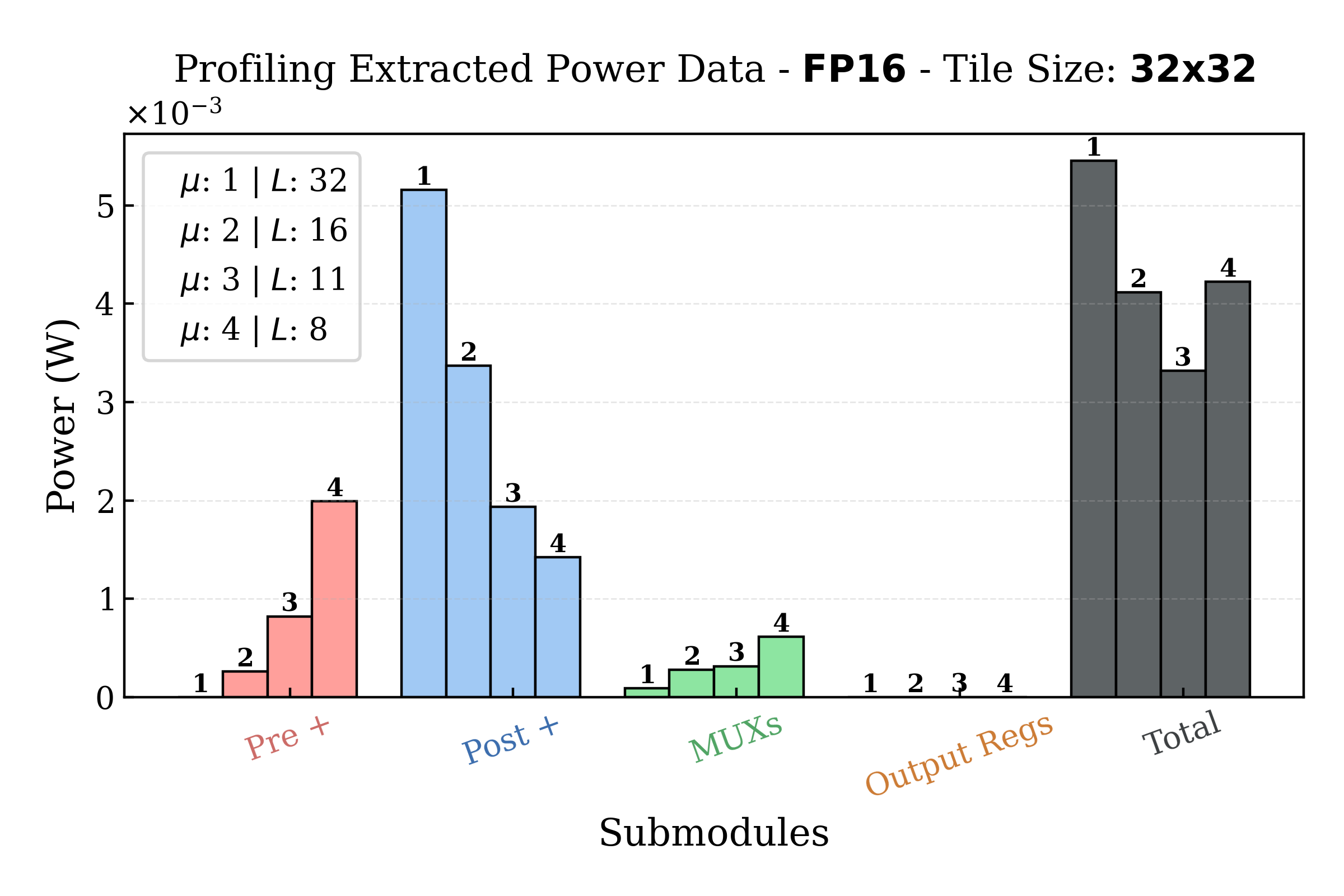}
        \vspace{-6mm}
        \caption{VCD-annotated synthesis power results.}
        \label{fig:surface_plot_fp}
    \end{subfigure}
    \caption{Area and power breakdown of the $32 \times 32$ configurations with FP16 adders and variable group size. Both metrics exhibit the same minima at optimal $\mu=3$.}
    \label{fig:profiling_results}
\end{figure}

\subsection{Submodule Cost Breakdown}\label{sec:histogram}

To assess the relative contributions of the submodules defined in Figure~\ref{fig:arch_overview}, we synthesized a $32 \times 32$ tile configuration using FP16 adders. We measured both area and average power consumption (via VCD-annotated simulations). The results, presented in Figure~\ref{fig:profiling_results}, empirically confirm the trade-off predicted by the analytical model.
As the group size $\mu$ increases, the cost of the \textit{Accumulate+} logic (orange) decreases linearly, while the cost of the \textit{Build+} logic (blue) grows exponentially. For this specific configuration ($32\times32$, FP16), the minimum total area is achieved at $\mu = 3$.  Since the simulated power exhibits the same scaling trends and optima as area, the remainder of this analysis focuses on area as the primary optimization metric.

\subsection{Validation of Analytical Model}
\begin{figure}[!tb]
    \centering
    \begin{subfigure}[b]{0.45\textwidth}
        \includegraphics[width=\textwidth]{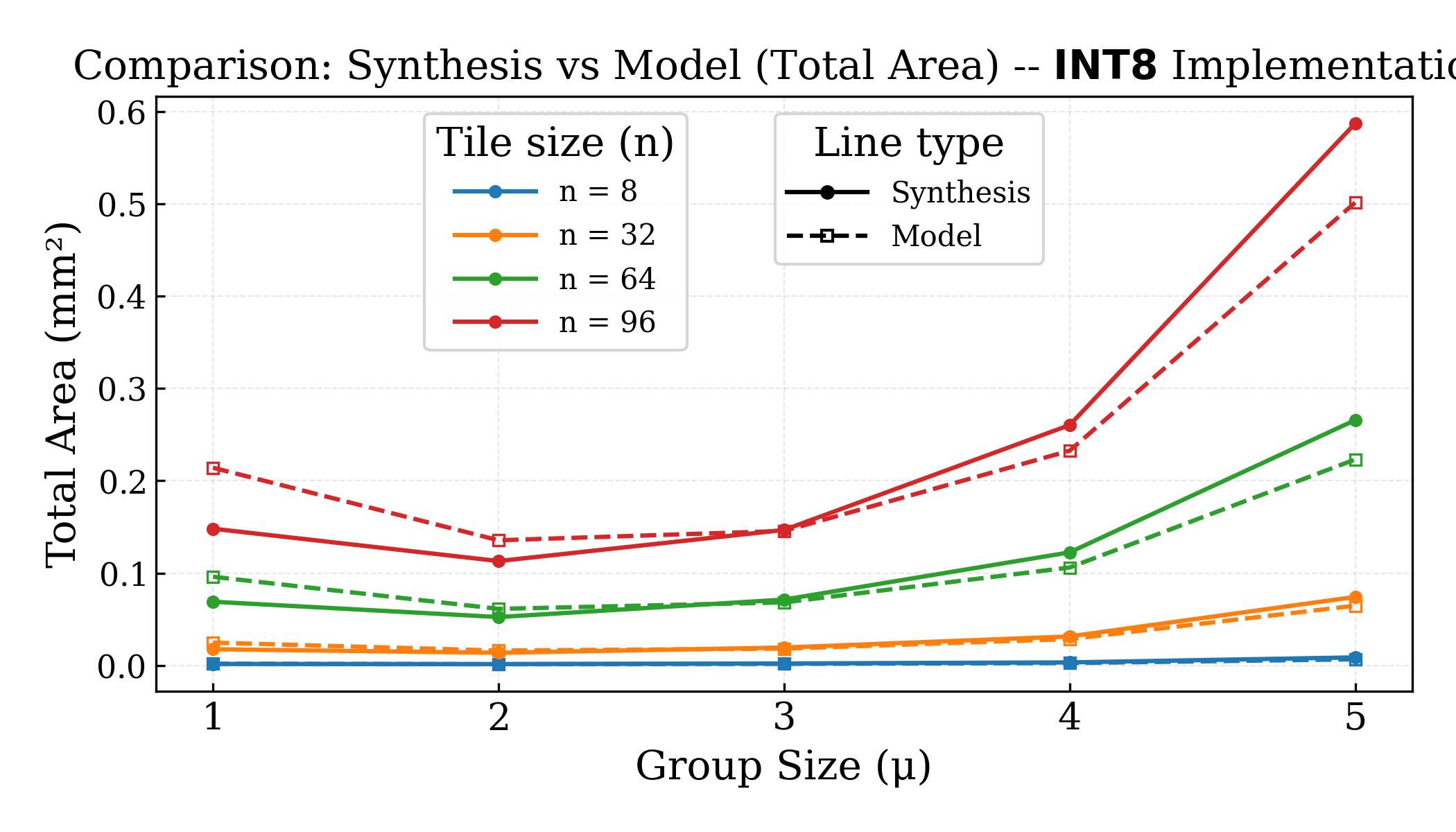}
        \caption{INT8 ($\gamma = 0.55$)}
        \label{fig:syn-results-int}
    \end{subfigure}
    \hfill
    \begin{subfigure}[b]{0.45\textwidth}
        \includegraphics[width=\textwidth]{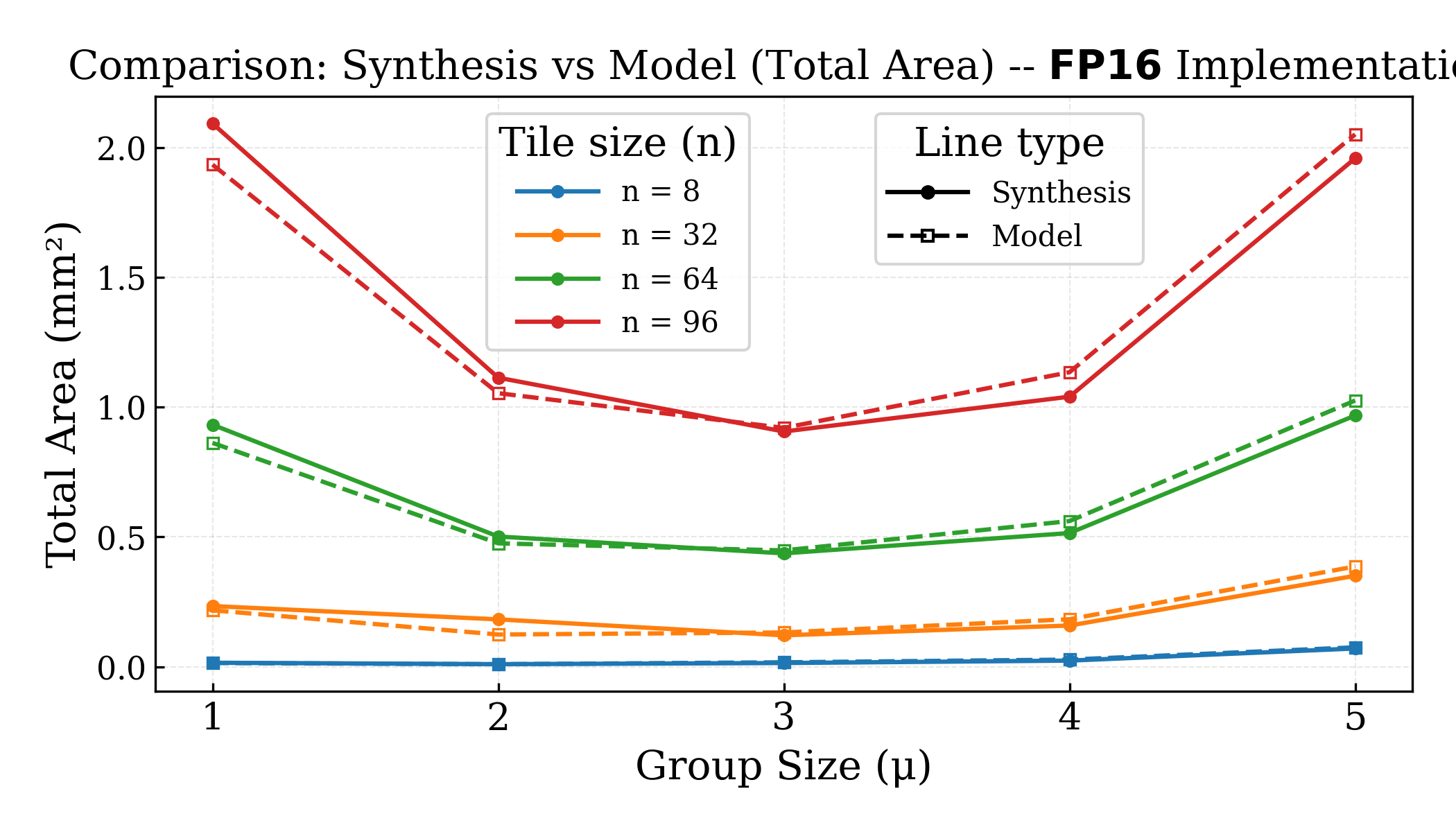}
        \caption{FP16 ($\gamma = 1.5$)}
        \label{fig:syn-results-fp}
    \end{subfigure}
    \caption{Comparison of analytical model predictions with synthesized area results across multiple design points. The model tracks hardware cost closely, validating its use for early-stage exploration. 
    }
    \label{fig:syn-results}
\end{figure}

To validate the model end-to-end, we generated and synthesized Verilog code for every design point listed in Table~\ref{tab:design_space}. The model coefficients $a_x$ were calibrated by synthesizing isolated unit cells (2-to-1 MUX, register, INT8/FP16 adder) in the target technology.

To account for implementation-dependent overheads not captured by the functional unit counts (e.g., control logic, buffering, and technology-specific optimizations) we apply a single, data-type specific global scaling factor ($\gamma$). This factor, determined by minimizing the MSE between the model and synthesis results, absorbs the linear overhead associated with the complexity of the data path. Figure~\ref{fig:syn-results} compares the analytical predictions against the synthesis data. The model tracks the synthesis results with high fidelity across all tile sizes and group sizes. This confirms that the derived scaling equations (Eq.~\ref{eq:scale_pre}-\ref{eq:scale_reg}) accurately capture the architectural behavior, validating the model's utility for rapid, pre-synthesis exploration.

\subsection{Impact of Activation Data Type}
The cost model shows that when multiplexers and inverters are comparable in size to adders, LUT-based designs offer little advantage over the baseline add-and-accumulate architecture. This case corresponds to $\mu = 1$ and is shown in Figure~\ref{fig:paper-overview} (middle). 

Here, the cost of additional read-out port (MUX and inverter) outweighs the benefits of reduced adder complexity. Conversely, when adders become significantly more expensive than MUX/inverter logic, LUT-based approaches become increasingly advantageous, and the optimal group size shifts toward larger $\mu$ values to minimize the $\text{Accumulate+}_\text{cost}$ term.

In short, our hypothesis is as follows:  

\begin{center}
\begin{tabular}{ c|c }
Adder-to-read-out Cost Ratio & LUT benefit \\
\hline
Balanced: $a_{add} \approx (a_{mux} + a_{inv})$ e.g., INT8 & Minimal \\
High: $a_{add} \gg (a_{mux} + a_{inv})$ e.g., FP16 & Significant \\
\end{tabular}
\vspace{2mm}
\end{center}

\noindent This hypothesis is confirmed in Figure~\ref{fig:syn-results}. For INT8, the area difference between the $\mu = 1$ case and optimal group size is negligible. In contrast, for FP16, a large improvement can be made by selecting $\mu > 1$.

\begin{figure}[!tb]
    \centerline{\includegraphics[width=0.99\linewidth]{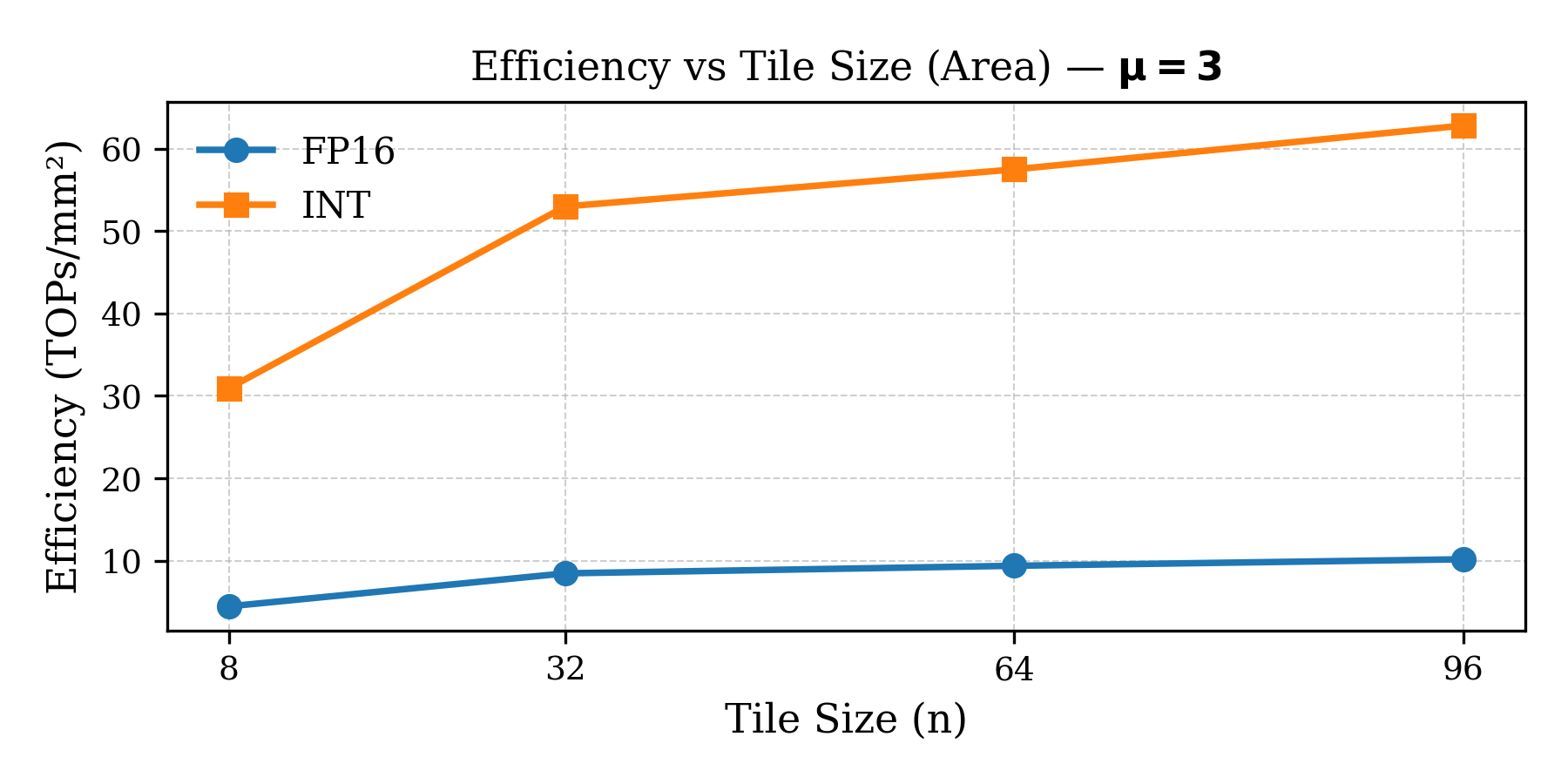}}
    \caption{Effect of tile size on area efficiency of LUT architecture with square tiles ($m=n$) optimal $\mu$ and FP16 activations. A higher tile size results in a higher throughput ($m \cdot n$) and total area. 
    }
    \label{fig:tops_per_mmsq}
\end{figure}

\begin{figure*}[!t]
    \centering
    \begin{subfigure}{0.95\linewidth}
        \centering
        \includegraphics[width=\linewidth]{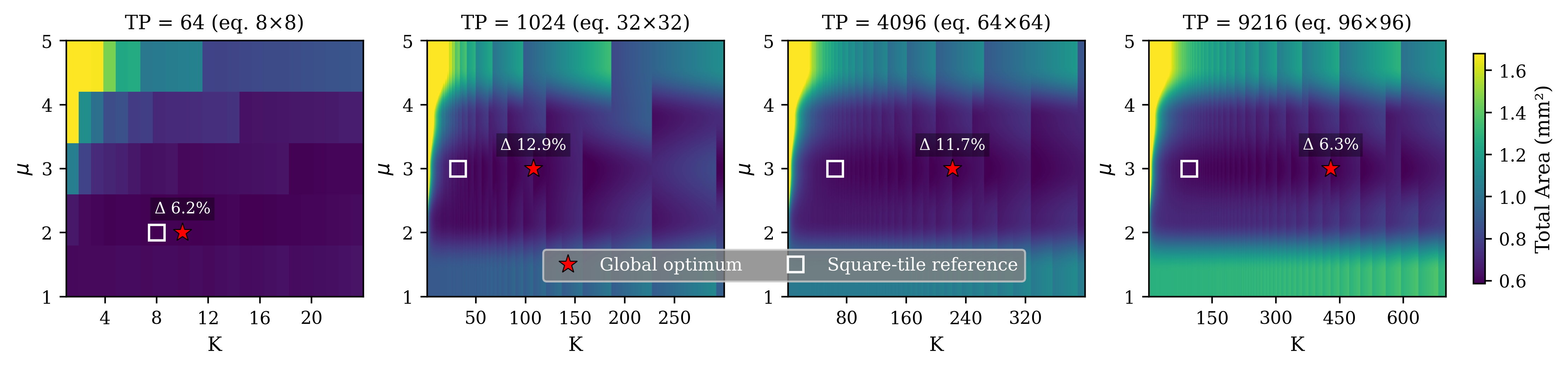}
        \vspace{-6mm}
        \caption{FP16 activations}
        \label{fig:elongated_tile_fp}
    \end{subfigure}
    \vspace{4mm}
    \begin{subfigure}{0.95\linewidth}
        \centering
        \includegraphics[width=\linewidth]{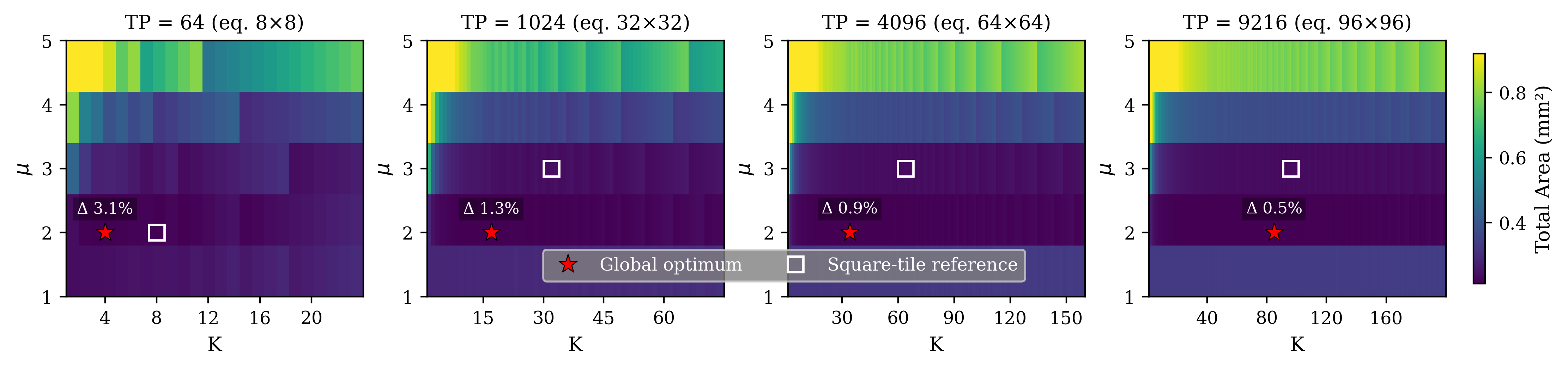}
        \vspace{-6mm}
        \caption{INT8 activations}
        \label{fig:elongated_tile_int}
    \end{subfigure}
    \caption{Effect of instantiating non-square tiles on area efficiency.
    $\Delta$ gives the relative area decrease compared to a square tile with optimal $\mu$.
    For FP16 activations, elongated tiles with more read-outs per LUT than total LUT inputs ($K > L \cdot \mu$) achieve lower area.
    For INT8 activations, the opposite trend holds, favoring tiles with more LUTs and fewer read-outs per LUT ($K < L \cdot \mu$).
    }
    \label{fig:elongated_tile}
\end{figure*}

\section{Design Space Exploration (DSE)}\label{sec:dse}
\subsection{Comparison to Baseline Architecture}\label{sec:baseline}

To assess the effectiveness of LUT-based architectures for ternary matrix multiplications in general, we compare our implementation against the two baselines illustrated in Figure~\ref{fig:paper-overview}. In the full-width multiplication baseline, ternary weights are first dequantized to FP16 and then processed by an FP16 multiplier. In the sign-flip multiplication baseline, the multiplier is replaced by a MUX that selects between the input, its inverse, or 0, depending on the ternary weight. 

Both baselines were implemented using standard design practices and synthesized for a $32\times32$ tile configuration. As summarized in Table~\ref{tab:area_comparison},  our optimized LUT-based design achieves a $2.23\times$ area reduction compared to the dequantization approach and a $1.64\times$ reduction compared to the optimized sign-flip baseline. This further confirms that for complex data types (FP16), a LUT-based approach yields substantial savings compared to a dedicated sign-flip unit.

\subsection{Scalability and Core Density}\label{sec:tops_per_mm}
Throughput scaling can be achieved by instantiating many small cores or fewer large cores. To evaluate which choice is preferred for a certain throughput target, we evaluated the area efficiency (TOPS/mm$^2$) across a range of tile sizes ($8\times8$ to $96\times96$). As illustrated in Figure~\ref{fig:tops_per_mmsq}, area efficiency improves monotonically with tile size. 

This trend is explained directly by the area equations derived in Section~\ref{sec:cost_model}. By dividing the total area (Eq.~\ref{eq:total_hw_cost}) by the throughput ($\sim n \cdot m$), we obtain the area cost per unit of throughput:

\begin{align}
    \begin{split}
        \frac{\text{Area}}{n \cdot m} \sim & \ a_{add} \left( \frac{1}{m} \cdot \frac{3.069^\mu}{1.938 \mu} + \frac{1}{\mu} \right) \\
        & + (a_{mux} + a_{inv}) \frac{3^\mu-1}{2\mu} + a_{reg} \frac{1}{n}
    \end{split}
\end{align}

This expression reveals the asymptotic behavior of the architecture. Accumulate ($\frac{1}{\mu}$) and MUX ($\frac{3^\mu-1}{2\mu}$) costs are independent of tile size. They represent the fundamental cost of the compute logic. In contrast, the Build cost is proportional to $\frac{1}{m}$, and the Output Register cost is proportional to $\frac{1}{n}$. As the tile dimensions $n$ and $m$ increase, these overhead terms vanish, leaving only the fundamental compute cost. Consequently, maximizing core size is the optimal strategy for area efficiency.

\begin{table}[]
    \centering
    \renewcommand{\arraystretch}{1.2} 
    \caption{Area comparison between baselines and the LUT-based design ($32\times32$ tile, FP16).}
    \begin{tabular}{l c c}
    \hline
    \textbf{Design} & \textbf{Area [mm$^2$]} & \textbf{Relative cost} \\
    \hline
    Full-width multiplication baseline       & 0.268 &\textbf{2.23$\times$} \\
    Sign-flip multiplication baseline & 0.197 & \textbf{1.64$\times$} \\
    \makecell[l]{\textbf{This work} (optimal $\mu=3$)} & \textbf{0.120} & 1$\times$ \\
    \hline
    \end{tabular}
    \label{tab:area_comparison}
\end{table}

\begin{table*}[h]
\centering
\caption{Area comparison with state-of-the-art implementations. The reported numbers only include a single-core instantiation. \\
\small $^*$Number likely includes buffers of unknown size. \\
$\dagger$Using scaling factor of 0.62 for delay and 0.41 for area~\cite{scaling}.
}

\begin{tabular}{l c c c c c c c c c c c}
\hline
\textbf{Work} & \textbf{\makecell{Throughput \\  {[}mul/cycle{]}}} & \textbf{L} & \textbf{$\mu$} & \textbf{K} & \textbf{Weight} & \textbf{Act. type} & \textbf{Tech} & \textbf{\makecell{\textbf{$f_{clk}$} \\ {[}MHz{]}}} & \textbf{Core area} & \makecell{\textbf{Area} \\ \textbf{decrease}} & \makecell{\textbf{Model} \\ \textbf{prediction}} \\ 
\hline
TENET~\cite{tenet} & 2048 & 32 & 2 & 32 & 1.6b & INT8 & 28nm & 500  & 640 000$^*$ $\mu m^2$ &  \multirow{2}{*}{7.9$^{\dagger}\times$} & \multirow{2}{*}{1.004$\times$}  \\
Ours              & 2040 & 34 & 2 & 30 & 1.6b & INT8 & 16nm & 800$^{\dagger}$ & 33 125 $\mu m^2$  &  \\
\hline
TeLLMe v2~\cite{tellmev2} & 1344 & 28 & 3 & 16 & 1.6b & INT8 & \multirow{2}{*}{\makecell{Zynq UltraScale+\\ XCK26 MPSoC}} & \multirow{2}{*}{\makecell{N/A}} & 35 200 LUT & \multirow{2}{*}{\makecell{$1.4\times$}} & \multirow{2}{*}{1.22$\times$} \\
Ours                        & 1334 & 26 & 2 & 23 & 1.6b & INT8 & & & 25 709 LUT &   \\
\hline
\end{tabular}

\label{tab:comparison_work_synth}
\end{table*}

\subsection{Optimal Tile Geometry}\label{sec:elongated}
While the synthesis results in Section~\ref{sec:validate} assumed square tiles ($m=n$), our analytical model allows us to explore rectangular geometries ($m \neq n$) to further optimize efficiency. We investigate the optimal aspect ratio for both INT8 and FP16 data types, representing balanced and high adder-to-read-out cost ratios, respectively.

Figure~\ref{fig:elongated_tile} reveals that the area-optimal architecture consistently deviates from a square tile and is strictly governed by the activation data type. For FP16 activations, the high cost of adders discourages LUT replication ($L$), while sign inversion is inexpensive (a simple bit-flip). This shifts the optimal design toward maximizing the reuse of each LUT entry across many parallel read-out ports ($K$), favoring elongated tiles where $K > L \cdot \mu$. Conversely, for INT8 activations, sign inversion requires a dedicated adder for two's complement operations, increasing the relative cost of the read-out logic. Consequently, it becomes more area-efficient to replicate the cheaper LUTs ($L$) rather than widening the expensive fetch units ($K$). This favors tiles where $L \cdot \mu > K$.

\section{Results and State-of-the-Art Comparison} \label{sec:results}

In this section, we revisit the related works to make a quantitative comparison. For two works that have reported isolated area numbers of the LUT-based GEMV core, we used our analytical model to determine the area-optimal $n$, $K$ and $\mu$ values required to achieve comparable throughput. We then instantiated these optimal configurations using our hardware generator and synthesized them to compare the area footprint under identical constraints. Table~\ref{tab:comparison_work_synth} shows the results, proving that existing works use suboptimal architectural parameters. We estimate that up to 22\% improvement can be achieved by choosing better design parameters, but potentially much more by making use of the more efficient implementation of our generator.

A quantitative comparison with other works was not possible due to a lack of reported area breakdown, but our model further indicates that design choices are often suboptimal. For example, Slim-LLaMA~\cite{kim_slim_lama} opted for many small LUT cores with limited read-outs per LUT and low-precision INT activations. However, our results strongly indicate that such design choices are unlikely to lie on the true efficiency frontier.

\FloatBarrier

\section{Conclusion}
This work establishes a systematic framework to explore the design space of ternary LUT-based accelerators, an increasingly relevant direction for efficient LLM inference, and addresses inconsistencies across prior designs. By combining an open-source RTL generator with an analytical model validated through synthesis, we enable exploration of the full architectural space and uncover trends that challenge several assumptions in recent literature. Specifically, we find that: 1) LUT-based architectures yield diminishing returns for small activation data types; 2) for fixed throughput, a few large LUT cores are significantly more area-efficient than many small ones; and 3) the optimal architectural parameters are strongly determined by the activation data type.

These insights demonstrate that substantial area savings are achievable under matched throughput, but only when designs adopt suitable tile shapes, sufficient LUT fan-out, and activation types that justify ternary-weight arithmetic. Our evaluation reveals that prior accelerators often deviate from these optimal points, explaining the $1.4\times$ architectural improvement we demonstrate over a state-of-the-art design. With a publicly available generator and reproducible methodology, this work establishes a common basis for fair comparison and offers concrete guidance for the design of future ternary-weight inference accelerators.
\section{Acknowledgments}
This project has been partly funded by the European Research Council (ERC) under grant agreement No. 101088865, the European Union’s Horizon 2020 program under grant agreement No. 101070374, the Flanders AI Research Program, Research Foundation Flanders (FWO) under grant No. 1S37125N, and KU Leuven.

\bibliographystyle{ieeetr}
\bibliography{references}

@online{wang_bitnet_2023,
	title = {{BitNet}: Scaling 1-bit Transformers for Large Language Models},
	titleaddon = {{arXiv}.org},
	author = {Wang, Hongyu and Ma, Shuming and Dong, Li},
	date = {2023-10-17},
	langid = {english},
}

@INPROCEEDINGS{mo_lut_2025,
  title      = "{LUT} Tensor Core: A {Software-Hardware} {Co-Design} for
                {LUT-Based} {Low-Bit} {LLM} Inference",
  booktitle  = "Proceedings of the 52nd Annual International Symposium on
                Computer Architecture",
  author     = "Mo, Zhiwen and Wang, Lei and Wei, Jianyu and Zeng, Zhichen and
                Cao, Shijie and Ma, Lingxiao and Jing, Naifeng and Cao, Ting
                and Xue, Jilong and Yang, Fan and Yang, Mao",
  publisher  = "ACM",
  pages      = "514--528",
  month      =  jun,
  year       =  2025,
  address    = "New York, NY, USA",
  conference = "ISCA '25: Proceedings of the 52nd Annual International
                Symposium on Computer Architecture",
  location   = "Tokyo Japan"
}

@misc{park_lut-gemm_2024,
	title = {{LUT}-{GEMM}: Quantized Matrix Multiplication based on {LUTs} for Efficient Inference in Large-Scale Generative Language Models},
	number = {{arXiv}:2206.09557},
	publisher = {{arXiv}},
	author = {Park, Gunho and Park, Baeseong and Kim, Minsub},
	date = {2024-04-01},
	eprinttype = {arXiv},
	eprint = {2206.09557},
	note = {arXiv:2206.09557},
}

@INPROCEEDINGS{jeon_biqgemm_2020,
  title           = "{BiQGEMM}: Matrix multiplication with lookup table for
                     binary-coding-based quantized {DNNs}",
  booktitle       = "{SC20}: International Conference for High Performance
                     Computing, Networking, Storage and Analysis",
  author          = "Jeon, Yongkweon and Park, Baeseong and Kwon, Se Jung and
                     Kim, Byeongwook and Yun, Jeongin and Lee, Dongsoo",
  publisher       = "IEEE",
  month           =  nov,
  year            =  2020,
  copyright       = "https://ieeexplore.ieee.org/Xplorehelp/downloads/license-information/IEEE.html",
  conference      = "SC20: International Conference for High Performance
                     Computing, Networking, Storage and Analysis",
  location        = "Atlanta, GA, USA"
}

@INPROCEEDINGS{park_figlut_2025,
  title           = "{FIGLUT}: An energy-efficient accelerator design for
                     {FP-INT} {GEMM} using look-up tables",
  booktitle       = "2025 {IEEE} International Symposium on High Performance
                     Computer Architecture ({HPCA})",
  author          = "Park, Gunho and Kwon, Hyeokjun and Kim, Jiwoo and Bae,
                     Jeongin and Park, Baeseong and Lee, Dongsoo and Lee,
                     Youngjoo",
  publisher       = "IEEE",
  pages           = "1098--1111",
  month           =  mar,
  year            =  2025,
  conference      = "2025 IEEE International Symposium on High Performance
                     Computer Architecture (HPCA)",
  location        = "Las Vegas, NV, USA"
}

@INPROCEEDINGS{kim_slim_lama,
  author={Kim, Sangyeob and Lee, Jungwan and Yoo, Hoi-Jun},
  booktitle={2025 IEEE International Solid-State Circuits Conference (ISSCC)}, 
  title={{Slim-Llama: A 4.69mW Large-Language-Model Processor with Binary/Ternary Weights for Billion-Parameter Llama Model}}, 
  year={2025},
  volume={68},
  number={},
  pages={421-423},
  doi={10.1109/ISSCC49661.2025.10904761}
}

@misc{qiao_tellme_2025,
	title = {{TeLLMe}: {An} {Energy}-{Efficient} {Ternary} {LLM} {Accelerator} for {Prefilling} and {Decoding} on {Edge} {FPGAs}},
	url = {http://arXiv.org/abs/2504.16266},
	doi = {10.48550/arXiv.2504.16266},
	urldate = {2025-04-27},
	publisher = {arXiv},
	author = {Qiao, Ye and Cheng, Zhiheng and Zhang, Yifan and Wang, Yian and Huang, Sitao},
	month = apr,
	year = {2025},
	note = {arXiv:2504.16266 [cs]},
}

@article{JMLR:v26:24-2050,
  author  = {Hongyu Wang and Shuming Ma and Lingxiao Ma and Lei Wang and Wenhui Wang and Li Dong and Shaohan Huang and Huaijie Wang and Jilong Xue and Ruiping Wang and Yi Wu and Furu Wei},
  title   = {{BitNet: 1-bit Pre-training for Large Language Models}},
  journal = {Journal of Machine Learning Research},
  year    = {2025},
  volume  = {26},
  number  = {125},
  pages   = {1--29},
  url     = {http://jmlr.org/papers/v26/24-2050.html}
}

@misc{chen2024,
      title={{TernaryLLM: Ternarized Large Language Model}},
      author={Tianqi Chen and Zhe Li and Weixiang Xu and Zeyu Zhu and Dong Li and Lu Tian and Emad Barsoum and Peisong Wang and Jian Cheng},
      year={2024},
      eprint={2406.07177},
      archivePrefix={arXiv},
      primaryClass={cs.LG},
      url={https://arXiv.org/abs/2406.07177}, 
note={arXiv:2406.07177}
}

@misc{kaushal2024spectrasurprisingeffectivenesspretraining,
      title={{Spectra: Surprising Effectiveness of Pretraining Ternary Language Models at Scale}}, 
      author={Ayush Kaushal and Tejas Vaidhya and Arnab Kumar Mondal and Tejas Pandey and Aaryan Bhagat and Irina Rish},
      year={2024},
      eprint={2407.12327},
      archivePrefix={arXiv},
      primaryClass={cs.LG},
      url={https://arXiv.org/abs/2407.12327}, 
note={arXiv:2407.12327}
}

@misc{sundaram2024llavaolmobitnet1bternaryllmgoes,
      title={{LLaVaOLMoBitnet1B: Ternary LLM goes Multimodal!}}, 
      author={Jainaveen Sundaram and Ravi Iyer},
      year={2024},
      eprint={2408.13402},
      archivePrefix={arXiv},
      primaryClass={cs.LG},
      url={https://arXiv.org/abs/2408.13402}, 
note={arXiv:2408.13402}
}

@misc{badshah2024quantifyingcapabilitiesllmsscale,
      title={{Quantifying the Capabilities of LLMs across Scale and Precision}}, 
      author={Sher Badshah and Hassan Sajjad},
      year={2024},
      note={arXiv:2405.03146},
      archivePrefix={arXiv},
      primaryClass={cs.LG},
      url={https://arXiv.org/abs/2405.03146},
}

@misc{ma2024era1bitllmslarge,
      title={{The Era of 1-bit LLMs: All Large Language Models are in 1.58 Bits}}, 
      author={Shuming Ma and Hongyu Wang and Lingxiao Ma and Lei Wang and Wenhui Wang and Shaohan Huang and Li Dong and Ruiping Wang and Jilong Xue and Furu Wei},
      year={2024},
      eprint={2402.17764},
      archivePrefix={arXiv},
      primaryClass={cs.CL},
      url={https://arXiv.org/abs/2402.17764}, 
}

@misc{chisel-float,
  author       = {Robin Geens},
  title        = {chisel-float: Mixed-precision floating point units, wrapped in Chisel},
  howpublished = {\url{https://github.com/KULeuven-MICAS/chisel-float}},
  year         = {2025},
}

@misc{wang2025bitnetv2native4bit,
      title={{BitNet v2: Native 4-bit Activations with Hadamard Transformation for 1-bit LLMs}}, 
      author={Hongyu Wang and Shuming Ma and Furu Wei},
      year={2025},
      eprint={2504.18415},
      archivePrefix={arXiv},
      primaryClass={cs.CL},
      url={https://arxiv.org/abs/2504.18415}, 
}

@misc{tenet,
      title={{TENET: An Efficient Sparsity-Aware LUT-Centric Architecture for Ternary LLM Inference On Edge}}, 
      author={Zhirui Huang and Rui Ma and Shijie Cao and Ran Shu and Ian Wang and Ting Cao and Chixiao Chen and Yongqiang Xiong},
      year={2025},
      eprint={2509.13765},
      archivePrefix={arXiv},
      primaryClass={cs.AR},
      url={https://arxiv.org/abs/2509.13765}, 
}

@misc{tellmev2,
      title={TeLLMe v2: An Efficient End-to-End Ternary LLM Prefill and Decode Accelerator with Table-Lookup Matmul on Edge FPGAs}, 
      author={Ye Qiao and Zhiheng Chen and Yifan Zhang and Yian Wang and Sitao Huang},
      year={2025},
      eprint={2510.15926},
      archivePrefix={arXiv},
      primaryClass={cs.AR},
      url={https://arxiv.org/abs/2510.15926}, 
}

@article{scaling,
title = {Scaling equations for the accurate prediction of CMOS device performance from 180nm to 7nm},
journal = {Integration},
volume = {58},
pages = {74-81},
year = {2017},
issn = {0167-9260},
doi = {https://doi.org/10.1016/j.vlsi.2017.02.002},
url = {https://www.sciencedirect.com/science/article/pii/S0167926017300755},
author = {Aaron Stillmaker and Bevan Baas}
}

\end{document}